\documentclass[alpha-refs]{wiley-article}

\usepackage{graphicx}
\usepackage[space]{grffile}
\usepackage{subcaption}
\usepackage{latexsym}
\usepackage{textcomp}
\usepackage{longtable}
\usepackage{tabulary}
\usepackage{booktabs,array,multirow}
\usepackage{amsfonts,amsmath,amssymb}
\usepackage{enumitem}
\usepackage{natbib}
\usepackage{url}
\usepackage{hyperref}
\usepackage[capitalise,noabbrev]{cleveref}
\usepackage[dvipsnames]{xcolor}
\hypersetup{colorlinks=true,linkcolor=blue,citecolor=blue,urlcolor=blue}
\usepackage{etoolbox}
\usepackage[normalem]{ulem}

\AtBeginDocument{\DeclareGraphicsExtensions{.pdf,.PDF,.eps,.EPS,.png,.PNG,.tif,.TIF,.jpg,.JPG,.jpeg,.JPEG}}
\usepackage[utf8]{inputenc}
\usepackage[english]{babel}

\usepackage{listings}
\definecolor{codebg}{RGB}{248,248,248}
\definecolor{codeframe}{RGB}{200,200,200}
\definecolor{keyword}{RGB}{0,0,180}
\definecolor{comment}{RGB}{0,128,0}
\definecolor{string}{RGB}{163,21,21}
\lstdefinelanguage{SysMLv2}{
  keywords={package, part, port, item, attribute, connection, flow, state,
            action, requirement, constraint, verification, analysis, calc,
            view, viewpoint, concern, stakeholder, actor, subject, enum,
            abstract, private, public, import, def, exhibit, perform, first,
            then, transition, accept, connect, to, from, in, out, doc,
            variation, variant, allocation, allocate, assume, require,
            satisfy, verify, objective, frame, metadata, return, use, case,
            ref, end, redefines, subsets, expose, render, filter, by},
  keywordstyle=\color{keyword}\bfseries,
  sensitive=true,
  morecomment=[l]{//},
  morecomment=[s]{/*}{*/},
  morestring=[b]",
  commentstyle=\color{comment}\itshape,
  stringstyle=\color{string},
}
\crefname{lstlisting}{Listing}{Listings}
\Crefname{lstlisting}{Listing}{Listings}
\crefname{theorem}{Theorem}{Theorems}
\crefname{proposition}{Proposition}{Propositions}
\crefname{lemma}{Lemma}{Lemmas}
\crefname{corollary}{Corollary}{Corollaries}
\crefname{definition}{Definition}{Definitions}

\newtheorem{definition}[theorem]{Definition}
\newtheorem{remark}[theorem]{Remark}

\newcommand{\Beh}{\mathcal{B}}
\newcommand{\C}{\mathcal{C}}
\newcommand{\Csys}{\mathcal{C}_{\mathrm{sys}}}
\newcommand{\Csib}{\mathcal{C}_{\mathrm{sib}}}
\newcommand{\Cq}{\mathcal{C}_{\mathrm{q}}}
\newcommand{\Cdisp}{\mathcal{C}_{\mathrm{disp}}}
\newcommand{\Cnaive}{\mathcal{C}_{\mathrm{naive}}}
\newcommand{\A}{A}
\newcommand{\G}{G}
\newcommand{\Env}{\mathcal{E}}
\newcommand{\Imp}{\mathcal{M}}
\newcommand{\refines}{\preccurlyeq}
\newcommand{\comp}{\otimes}
\newcommand{\quotient}{/}

\newcommand{\stl}{STL}
\newcommand{\rob}{\rho}
\newcommand{\Until}{\,\mathcal{U}\,}
\newcommand{\Ev}{\Diamond}
\newcommand{\Alw}{\Box}
\newcommand{\sig}{w}
\newcommand{\horizon}{T}

\newcommand{\riskeps}{\varepsilon}

\providecommand{\llbracket}{[\![}
\providecommand{\rrbracket}{]\!]}
\newcommand{\sysml}[1]{\texttt{#1}}

\makeatletter
\apptocmd{\wiley@affilmetadata}{\vskip10pt\textbf{Code availability}\\ C-BASE: \url{https://github.com/SEIL-ISR/cbase}}
\makeatother

\papertype{Original Article}
\paperfield{Systems Engineering}
\corraddress{Sai Sandeep Damera, Department of Electrical and Computer Engineering, University of Maryland, College Park, MD 20742, USA}
\corremail{sdamera@umd.edu}
\fundinginfo{This work was partially supported by the Lockheed Martin Chair in Systems Engineering and the NVIDIA Academic Grant Program using NVIDIA RTX PRO 6000 Blackwell GPUs and the NVIDIA DRIVE AGX Thor Developer Kit.}

\title{Contract-Based Agentic Systems Engineering:\\
Delegation as Quotient for Assured System  Design}
\author[1]{Sai Sandeep Damera}
\author[1]{Ryan Matheu}
\author[1]{John S.\ Baras}
\affil[1]{Department of Electrical and Computer Engineering and the Institute for Systems Research, University of Maryland, College Park, MD 20742, USA}
\runningauthor{Damera, Matheu, and Baras}

\makeatletter
\zref@newlabel{mdf@pagelabel-1}{\mdf@pagevalue{1}}
\makeatother
\begin{document}
\maketitle

\begin{abstract}
A component an agent builds from a prompt can satisfy it and still break the system it joins, because the prompt omits what its neighbours assume of it. The artifact to hand over is computed, not written. A SysML~v2 requirement with assumed and required constraints is an assume\slash guarantee contract, so when a system requirement is decomposed over fixed neighbours, the open component's requirement is the contract quotient, the weakest sound one, and again a SysML~v2 requirement. Its required constraints carry the neighbours' assumptions, which a requirement for the mission alone drops and unit tests with mocked neighbours cannot reach. Signal Temporal Logic is closed under the quotient. We build the requirements, with derivation, allocation, satisfaction and verification, as a SysML~v2 model and compute the quotient from it. A case study delegates the local controller of a simulated warehouse robot, whose route crosses a worker's path and ends at a dock, to one language model under the quotient requirement, a mission-only requirement and a natural-language task, and scores the deliveries on live episodes against the system requirement from ground truth.
\end{abstract}


\section{Introduction}
\label{cqd:sec:introduction}

\begin{figure}[!t]
\centering
\includegraphics[scale=0.93]{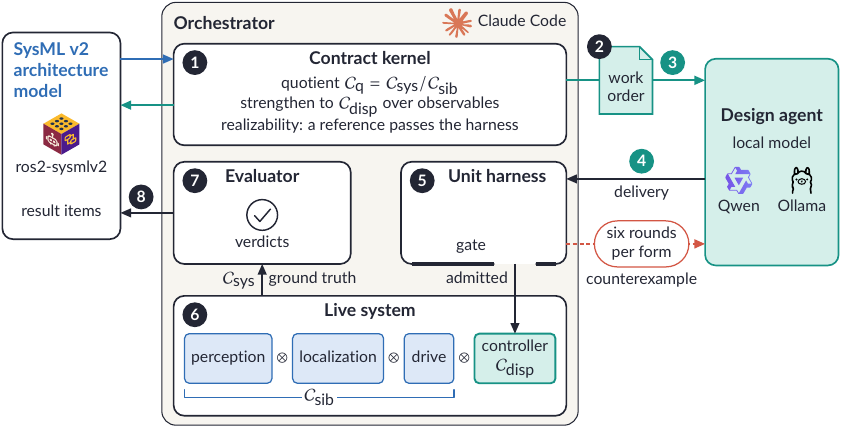}
\caption{The C-BASE architecture. The orchestrator agent is responsible for contact organization and delegation. A delegate contract is computed via the quotient operation and the unimplemented behavior is then shipped as a work order to the design agent. A candidate component is then subject to verification. Upon satisfaction of the local behaviors derived from the quotient contract, a candidate component is integrated into the system. If the overall system with the delegated component satisfies the system-level contract, it is accepted into the overall system design.}
\label{cqd:fig:orchestration}
\end{figure}

Requirement-driven model-based systems engineering (MBSE) establishes a framework for organizing a system's requirements and architecture in a single model. Requirements are then decomposed, allowing each system requirement to traverse down the architectural hierarchy to the components that must satisfy it. In SysML~v2, a requirement is a model element with a subject, assumed constraints, and required constraints, and the model relates it to the architecture and to the evidence by derivation, allocation, satisfaction, and verification \citep{omg2025sysml}. The components themselves are often designed by different parties, as when a prime contractor hands subsystems to suppliers under specifications and interface control documents. Each party receives a requirement that an engineer wrote by hand from the architecture, causing key missing functionality to be discovered at integration.

To elucidate the necessity of integration-aware constraints at the component level, we consider the development of a robot's autonomy stack. The perception component of a mobile robot in a warehouse environment transforms camera images and lidar scans into the positions of nearby people, localization estimates the robot's pose, a planner computes a feasible path,a local controller turns this into speed and turning commands, and an actuator executes them. We consider an autonomy stack with every component fixed except for the local controller, the implementation of which is delegated to a designer.

Each fixed component delivers assured output under an assumed condition on its environment. In closed-loop execution, such environmental properties are directly influenced by the behavior of the delegated component. In formal contract-based design, constraints on the operational environmental are referred to as the component's \emph{assumption}, and the output it then delivers under the premise that the assumptions are met is referred to as the \emph{guarantee} \citep{benveniste2008multiple,benveniste2018}. In the mobile robot autonomy stack, perception reports a person's position within its stated error only while the robot turns slowly enough for the camera image and the lidar scan to line up, and the actuation follows a command only below its designated saturation. Both assumptions therefore constrain the commands of the local controller. The requirements for the local controller rarely states these assumptions explicitly, so a controller can meet its local requirements and still break a neighbor's assumption, resulting in the system failing despite every component meeting its own specification. Unit tests that replace the neighbors with ideal stand-ins do not exercise such a failure, as an ideal stand-in has no assumption that the controller could break.

We consider designers that are generative agents, by which we mean a large language models that write a component's operational code or selects its design from a task statement and revises it on feedback from tests. Agent toolkits for robotics already let an agent choose a robot's tools and policies and let a language model write function bodies that automated checks accept or reject \citep{strandslabs2026}. Such an agent designs exactly against the prompt it is handed. What the agent learns in training carries no knowledge of the contracts of a particular architecture, such specifications human engineers learn in design reviews. It is also capable of delivering candidates faster than an integration team can investigate, so an omission in the task statement is more likely to reach integration.

We argue that the requirement handed to a designer should be computed from the architecture model. SysML~v2 reads a requirement as an implication, so that if all its assumed constraints hold, all its required constraints must hold \citep{omg2025sysml}. Contract theory calls this an assume/guarantee contract in saturated form, under which a component owes nothing when a behavior breaks its assumption. A contract \emph{refines} another when it accepts every environment the other accepts and promises no less. Delegation then poses the \emph{missing component problem} of contract-based design, which asks for a requirement on the missing component whose composition with the fixed components' contracts refines the system contract. We call such a requirement \emph{sound}. Every implementation of a sound requirement composes with the fixed components into a system that meets the system requirement in every environment it admits, so the architecture is correct by construction. The contract quotient \citep{incerQuotient2018} of the system contract by the fixed components' contracts is the most permissive sound requirement, so handing it over leaves the designer every implementation that any sound requirement allows. Its required constraints include the fixed components' assumptions, which we call \emph{preservation obligations} and which a requirement for the mission alone leaves out. Because the quotient is itself a SysML~v2 requirement, the requirement derived from it can replace the hand-written one in the architecture model, with a derivation from the system requirement and an allocation to the missing component. The contract theory we build on does not fix the formalism in which constraints are written \citep{benveniste2018}. To model robot behavior we write them in Signal Temporal Logic (\stl{}) \citep{malerNickovic2004}, which is closed under the quotient and whose robustness semantics \citep{fainekosPappas2009} scores every run by a signed margin. An engineer could write the neighbors' assumptions into a task statement by hand, but the quotient computes them and updates them when the neighbors' contracts change (\cref{cqd:sec:res-family}). \Cref{cqd:fig:orchestration} shows our framework for contract-based agentic systems engineering (C-BASE), which carries a delegation from the SysML~v2 architecture model through the quotient, a language model and a harness to live evidence that it writes back into the model.

\begin{figure}[!t]
\centering
\includegraphics{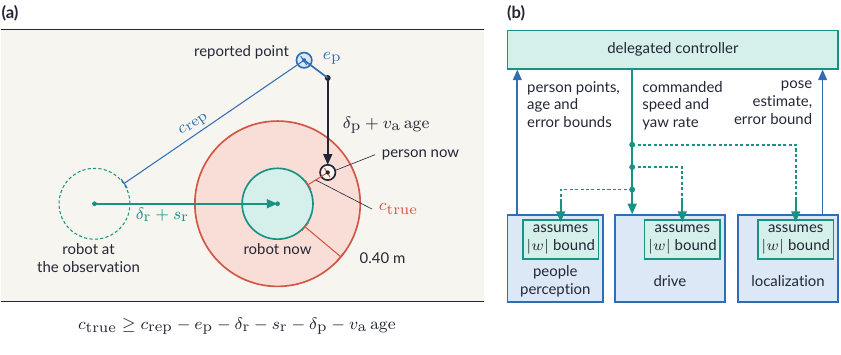}
\caption{(a) Since the most recent observation, reported with error $e_{\mathrm{p}}$, the robot has moved by up to $\delta_{\mathrm{r}} + s_{\mathrm{r}}$ and the person's nearest point has closed in by up to $\delta_{\mathrm{p}} + v_{\mathrm{a}}\,\mathrm{age}$, so the true clearance $c_{\mathrm{true}}$ can breach 0.40~m while the reported $c_{\mathrm{rep}}$ does not. (b) The controller relies on perception, localization and the drive, and each of them assumes a bound on the yaw rate the controller commands.}
\label{cqd:fig:coupling}
\end{figure}

The omission leads to integration failure in two ways. In the first, the delegated component treats a reported value as exact and current and so ignores a tolerance in a neighbor's guarantee. The engineer guards against this by strengthening the quotient into a requirement over the controller's observables (the signals it can see) that budgets every tolerance, and the contract kernel of C-BASE checks that this requirement refines the quotient. In the second, the delegated component breaks a neighbor's assumption, which the preservation obligations forbid. \Cref{cqd:fig:coupling} shows both mechanisms in our case study, in which a warehouse robot must cross a leg of open floor and rest at a dock while keeping 0.40~m from a crossing worker. The delegated local controller maps the pose, the path, the lidar scan and the person points to a speed and a yaw rate, while perception, localization and the drive stay fixed. We make three contributions.

\begin{enumerate}[leftmargin=*]
    \item A delegation theory in the requirement semantics of SysML~v2. Delegation poses the missing component problem, and the quotient of SysML~v2 requirements solves it with a SysML~v2 requirement that is the most permissive sound one, so the architecture is correct by construction. A requirement that omits the neighbors' assumptions is unsound because it admits implementations that break the system, and no unit-test harness with ideal stand-ins for the neighbors can tell these implementations apart from sound ones. A refinement lemma makes the strengthening onto observables sound, and a tick-by-tick reading of live episodes, which counts a failure against a component only while its assumption has held, relates live scoring to the theorems (\cref{cqd:sec:contracts,cqd:sec:delegation}).
    \item C-BASE holds a delegation's requirements in a SysML~v2 model. Its contract kernel computes the quotient and checks the engineer's strengthening onto observables. C-BASE then hands the result to a language model as its task, admits the deliveries in a unit-test harness with stand-ins for the neighbors and scores them live (\cref{cqd:sec:pillar}).
    \item A case study on a simulated warehouse robot whose local controller is delegated under the quotient-derived requirement, the mission-only requirement or a natural-language task. The fixed components are held to their contracts first, and the first failure of each live violation is attributed to its conjunct and owner (\cref{cqd:sec:case-study,cqd:sec:results}).
\end{enumerate}

\section{Related Work}
\label{cqd:sec:related-work}

Integrated Data-Driven and Model-Based Systems Engineering (IDDMBSE) extends the MBSE V-process for autonomous cyber-physical systems with a data-driven loop at every step \citep{baras2026iddmbseintegratingdatadrivenmodelbased}. Its tool chain maps SysML architectures to executable ROS autonomy stacks and combines formal, data-driven and runtime verification. When we reformulate IDDMBSE in SysML~v2, requirements become first-class model elements with assumed and required constraints which relates them to the architecture by derivation, allocation, satisfaction and verification \citep{omg2025sysml}. Syside is a SysML~v2 tool distribution from the vendor Sensmetry, with an editor and a modeler that validate models against the language's rules, a compiler that evaluates their expressions and a Python API to them \citep{vaicenavivcius2025syside}. The \texttt{ros2-sysmlv2} library defines the elements of a ROS~2 system in the language \citep{dameraAaC2026}. In this flow an engineer writes each component's requirement, and verification checks the design against it. We compute one component's requirement from the others' and keep it in the SysML~v2 model, derived from the system requirement.

Computing one component's requirement from the others' is the missing component problem, which goes back to submodule construction for communication protocols \citep{merlin1983construction}. Interface automata made a component's assumptions about its environment part of its interface \citep{dealfaro2001}, and assume/guarantee contracts divide the responsibility for a behavior between a component and its environment \citep{benveniste2008multiple}. AGREE checks that the contracts written for the components of an AADL architecture model compose to the contract of the system \citep{cofer2012}. The meta-theory of contracts of \citet{benveniste2018} treats assumptions and guarantees as assertions in a Boolean algebra. For assume/guarantee contracts the quotient solves the missing component problem, and \citet{incerQuotient2018} gave it in closed form for saturated contracts. Pacti computes the quotient for contracts whose assertions are polyhedral constraints \citep{pactiTCPS2025}.

Generative models now write both the architecture model and the components it describes. SysTemp generates SysML~v2 models from a natural-language specification with a team of language-model agents, and a parser's error report goes back to the writing agent until the model passes the grammar \citep{bouamra2025systempmultiagenttemplatebasedgeneration}. The generated model carries the requirements that the specification states, and the parser checks its syntax. Strands Robots builds a robot autonomy stack by agentic design, much as we do: an agent picks the robot tools and policies, and in the companion AI Functions a language model writes a function's body, while checks run on what the agent returns \citep{strandslabs2026}. Those checks are tool schemas, postconditions, success predicates over the simulator state and a model acting as judge, not assume/guarantee contracts with a composition and quotient calculus. No architecture model holds the components either, so nothing in the toolkits derives a component's requirement from the system's or checks that the neighbors' assumptions are preserved.
\section{Requirements as Contracts}
\label{cqd:sec:contracts}

\subsection{Assume/Guarantee Contracts}
\label{cqd:sec:ag-contracts}

We use the assume/guarantee contracts of \citet{benveniste2018}. A behavior is a signal $\sig$ that assigns a value to every system variable at every time in $[0, \horizon]$, and one episode of the running example is one behavior. An assertion is a subset of the set $\Beh$ of behaviors, with union, intersection and complement written $\cup$, $\cap$ and $\neg$. We identify a component $M$ with the assertion of the behaviors it can exhibit, so components compose by intersection.

\begin{definition}[refinement, composition, quotient]
\label{cqd:def:ops}
For saturated contracts $\C_1 = (\A_1, \G_1)$ and $\C_2 = (\A_2, \G_2)$, refinement is $\C_1 \refines \C_2$ iff $\A_1 \supseteq \A_2$ and $\G_1 \subseteq \G_2$; composition is $\C_1 \comp \C_2 = \big((\A_1 \cap \A_2) \cup \neg(\G_1 \cap \G_2),\; \G_1 \cap \G_2\big)$; and the quotient \citep{incerQuotient2018} is
\[
\C_1 \quotient \C_2 \;=\; \Big(\A_1 \cap \G_2,\;\; (\G_1 \cap \A_2) \cup \neg(\A_1 \cap \G_2)\Big).
\]
\end{definition}

\begin{proposition}[universal property of the quotient, \citealp{incerQuotient2018}]
\label{cqd:prop:adjoint}
For every saturated contract $\C$, $\C_2 \comp \C \refines \C_1$ if and only if $\C \refines \C_1 \quotient \C_2$, and $\C_1 \quotient \C_2$ is saturated. The quotient is therefore the largest contract whose composition with $\C_2$ refines $\C_1$.
\end{proposition}

\begin{proof}
Saturation holds because the guarantee contains the complement of the assumption. We prove the direction that \cref{cqd:thm:canonical} uses, from refinement of the quotient to refinement of the composition; the other direction is in \citet{incerQuotient2018}. Let $\C = (\A, \G) \refines \C_1 \quotient \C_2$ and write $\C_2 \comp \C = (\A_\comp, \G_\comp)$. We show $\G_\comp \subseteq \G_1$ and $\A_1 \subseteq \A_\comp$. First, $\G_\comp = \G_2 \cap \G \subseteq (\G_1 \cap \G_2 \cap \A_2) \cup (\G_2 \cap \neg\A_1) \subseteq \G_1$, the second inclusion by saturation of $\C_1$. Second, for a behavior $b \in \A_1$, either $b \notin \G_2 \cap \G$, so $b \in \A_\comp$, or $b \in \A_1 \cap \G_2 \subseteq \A$ and $b \in \G$, which forces $b \in \G_1 \cap \A_2$ and hence $b \in \A_2 \cap \A \subseteq \A_\comp$.
\end{proof}

\subsection{SysML~v2 Requirements}
\label{cqd:sec:sysml-reading}

A SysML~v2 requirement definition declares a subject, actors for the entities in its environment, and assumed and required constraints (clauses 7.21.1 and 7.21.2). ``The effective constraint for the requirement is then a logical implication: if all the assumption constraints are true, all the required constraints must be true'' (7.21.2, 8.4.17.1) \citep{omg2025sysml}. This implication holds on exactly the traces in $\varphi_\G \vee \neg\varphi_\A$, so a requirement definition with assumed constraints $\varphi_\A$ and required constraints $\varphi_\G$ denotes the saturated contract $(\varphi_\A, \varphi_\G \vee \neg\varphi_\A)$. In the architecture model of \cref{cqd:sec:pillar} the subject of every requirement definition is an episode trace (\cref{cqd:sec:scoring}), and the environment's behavior is stated in the assumed constraints. A satisfy relationship (7.21.4) claims that the episode trace it names lies in the saturated guarantee, and a verification case (7.24) records the evidence for that claim in its verdict. The claim $M \in \Imp(\C)$ quantifies over every episode $M$ can produce, so drawn episodes cannot establish it, and \cref{cqd:lem:binomial} bounds instead the probability that $M$ produces a violating episode under the operating distribution. SysML~v2 defines no composition or quotient of requirements. Its \sysml{Refinement} metadata marks a dependency without contract semantics (9.3.2.2.3), and specialization is not refinement, since a specialized requirement definition may add assumed constraints and so accept fewer environments. We therefore apply the operations of \cref{cqd:def:ops} to the contracts that requirements denote, and because every saturated contract is denoted by a requirement definition, the quotient of two requirements is again a requirement.

\subsection{Signal Temporal Logic as the Assertion Language}
\label{cqd:sec:stl}

Let $x(t)$ collect the real-valued variables of a behavior at time $t$. \stl{} formulas \citep{malerNickovic2004} are generated by $\varphi \Coloneqq \top \mid \mu_f \mid \neg\varphi \mid \varphi_1 \wedge \varphi_2 \mid \varphi_1 \Until_{[\tau_1,\tau_2]} \varphi_2$, where $\mu_f$ is the atomic predicate $f(x(t)) \geq 0$. The bounded until $\varphi_1 \Until_{[\tau_1,\tau_2]} \varphi_2$ holds at $t$ when $\varphi_2$ holds at some $t' \in [t+\tau_1, t+\tau_2]$ and $\varphi_1$ holds throughout $[t, t')$. Eventually is $\Ev_{[\tau_1,\tau_2]}\,\varphi = \top \Until_{[\tau_1,\tau_2]} \varphi$, and always is $\Alw_{[\tau_1,\tau_2]}\,\varphi = \neg\Ev_{[\tau_1,\tau_2]}\neg\varphi$. A formula denotes the set of signals that satisfy it at $t = 0$, with every interval intersected with $[0, \horizon]$. The robustness $\rob(\varphi, \sig, t)$ of \citet{fainekosPappas2009} is a signed margin: a positive robustness implies that $\sig$ satisfies $\varphi$ at $t$, and a negative one that it violates it. \Cref{cqd:sec:scoring} gives its form on sampled traces.

\begin{proposition}[closure]
\label{cqd:prop:closure}
If $\C_1$ and $\C_2$ are \stl{} contracts over a common signal space, then $\C_1 \comp \C_2$ and $\C_1 \quotient \C_2$ are \stl{} contracts, with formulas obtained by substituting the constituent formulas into \cref{cqd:def:ops}. Hence every requirement derived from \stl{} contracts by repeated composition and quotient is an \stl{} contract.
\end{proposition}

\begin{proof}
The operations of \cref{cqd:def:ops} are Boolean combinations of their operands, and the grammar is closed under negation and conjunction.
\end{proof}

\subsection{Sampled Fragment and Scoring Semantics}
\label{cqd:sec:scoring}

A trace samples one episode at its control ticks $t_0 < \cdots < t_N$, which in the case study are 0.05~s apart on average. Some signals at a tick are computed from a window of the record before it. The signal $s_{\mathrm{r}}$ is the age of the observation in hand times the largest speed commanded during that age (\cref{cqd:tab:notation}), and another is the largest $|w|$ over the 2.0~s before a stop. We treat these window signals as coordinates of $x$, so every atom is affine in $x$, although an atom that contains $s_{\mathrm{r}}$ is nonlinear in the recorded data. In the case study every declared contract (the system's and each fixed component's) assumes a safety formula $\Alw\, a$ and guarantees a conjunction of safety formulas $\Alw\, h_j$ with at most one bounded until $p \Until_{[0,\horizon]} q$. The formulas $a$ and $h_j$ are Boolean combinations of affine atoms, and $p$ and $q$ are conjunctions of affine atoms. The quotient is built from these formulas by \cref{cqd:def:ops}, and the dispatched requirement (the quotient strengthened onto the controller's observables, \cref{cqd:sec:dispatch}) keeps its assumption. That assumption contains, for each fixed component $i$, the implication $\alpha_i \rightarrow \gamma_i$ from its assumption $\alpha_i = \Alw\, a_i$ to its guarantee $\gamma_i = \Alw\, g_i$.

On a trace the robustness of an atom at $t_k$ is $f(x(t_k))$, negation takes the negative, conjunction and disjunction take the minimum and the maximum, $\Alw\, h$ takes the minimum over the ticks, and the bounded until has the robustness
\[
\rob(p \Until_{[0,\horizon]} q) \;=\; \max_{k \,:\, t_k \leq \horizon} \; \min\Big(\rob(q, t_k),\; \min_{j < k} \rob(p, t_j)\Big).
\]
Every atom is a non-strict inequality, so at robustness zero an atom holds while its negation, also at robustness zero, does not. The robustness therefore cannot decide a verdict at zero, and the kernel takes each verdict from the Boolean semantics and reports the robustness beside it. The declared conjuncts contain no negation, so for them a robustness of zero or more is equivalent to satisfaction. A contract holds when every conjunct does. Verdicts are read at the ticks.
To attribute a live failure, an integration engineer counts a guarantee's failure against its component only if the component's assumption held at every tick up to and including the tick that decides the failure. We score live verdicts in the same way.

\begin{definition}[prefix reading]
\label{cqd:def:prefix}
Let a contract assume $\psi = \Alw\, a \wedge \bigwedge_i (\Alw\, a_i \rightarrow \Alw\, g_i)$, where a declared contract has no implications. On a trace, $\psi$ holds so far at a tick if $a$ has held at every tick up to it and, for each $i$, $g_i$ has held at every tick up to it or $a_i$ has already failed. We write $\mathrm{H}\,\psi$ at a tick if $\psi$ has held so far at every tick up to and including it. A safety conjunct $\Alw\, h$ of the guarantee is decided violated at the first tick at which $h$ fails. An until $p \Until_{[0,\horizon]} q$ is decided satisfied at the first tick of its interval at which $q$ holds with $p$ held before. Otherwise it is decided violated at the first tick at which $p$ fails or at the end of its interval (the trace's last tick no later than $\horizon$). A conjunct decided violated is violated if $\mathrm{H}\,\psi$ holds at the deciding tick, same-tick if $\mathrm{H}\,\psi$ first fails there and vacuous if it failed earlier, and every other conjunct is satisfied. The contract's verdict is the most severe verdict among its conjuncts, in the order violated, same-tick, vacuous, satisfied.
\end{definition}

On the trace of a complete episode every conjunct is decided, and an until is pending only on the prefix of an episode still running. The rule for the implications follows the attribution of \cref{cqd:sec:evidence}.

\begin{proposition}[prefix reading refines the implication]
\label{cqd:prop:prefix}
Let a contract with guarantee formula $\theta$ assume a conjunction of safety formulas $\psi = \Alw\, a$. If its prefix verdict on a trace is not violated, the trace satisfies $\neg\psi \vee \theta$. If it is violated, the trace satisfies $\neg\psi \vee \theta$ exactly when $\psi$ fails, and then $\mathrm{H}\,\psi$ fails after the deciding tick.
\end{proposition}

\begin{proof}
If $\psi$ holds, $\mathrm{H}\,\psi$ holds at every tick, so a conjunct decided violated would make the verdict violated. When the verdict is not violated, no conjunct is decided violated: each safety conjunct holds at every tick and each until is decided satisfied, so $\theta$ holds. If $\psi$ fails, the implication holds vacuously. A violated verdict has a conjunct decided violated, so $\theta$ fails and the implication holds exactly when $\psi$ fails. Then $\mathrm{H}\,\psi$, which held at the deciding tick, fails at a later one.
\end{proof}

For the quotient and the dispatched requirement the prefix reading departs from the whole trace on one kind of trace. We call a trace excused if it satisfies $\psi$ and some $g_i$ fails at a tick at which $a_i$ has held so far.

\begin{theorem}[prefix reading of a delegated requirement]
\label{cqd:thm:prefix}
Let a contract with guarantee formula $\theta$ assume $\psi = \Alw\, a \wedge \bigwedge_i (\Alw\, a_i \rightarrow \Alw\, g_i)$, with every $\Alw\, a_i$ a conjunct of $\theta$. A violated prefix verdict relates to the whole trace as in \cref{cqd:prop:prefix}. A prefix verdict that is not violated is a satisfied verdict on the whole trace unless the trace is excused. On an excused trace the $a_i$ whose $g_i$ failed fails later, so the contract is violated on the whole trace, while the prefix reading counts no failure decided at or after the first tick at which a fixed component's guarantee fails with its assumption held so far.
\end{theorem}

\begin{proof}
At a tick at which $a_i$ has held so far, the $i$-th implication holds so far only if $g_i$ has held so far. So $\mathrm{H}\,\psi$ holds at every tick exactly when $a$ holds at every tick and each $g_i$ at every tick at which $a_i$ has held so far. Such a trace satisfies $\psi$, since under $\Alw\, a_i$ every tick is one at which $a_i$ has held so far. A trace that satisfies $\psi$ and is not excused is therefore one on which $\mathrm{H}\,\psi$ holds at every tick, and the proof of \cref{cqd:prop:prefix} applies to it, as it does to the violated verdicts and to the traces on which $\psi$ fails. On an excused trace $g_i$ fails, so $\psi$ forces $\Alw\, a_i$ to fail after the tick at which $g_i$ failed, and $\theta$, which contains $\Alw\, a_i$, fails while $\psi$ holds. There $\mathrm{H}\,\psi$ fails at the first tick at which some $g_j$ fails with $a_j$ held so far, and a conjunct decided violated from then on is same-tick or vacuous.
\end{proof}

\Cref{cqd:prop:prefix,cqd:thm:prefix} carry the theorems of \cref{cqd:sec:delegation} over to live evidence. The counts of violations under the two readings differ by the episodes whose assumption fails after a violation and, for the dispatched requirement, by the excused episodes without an earlier violation. The unit-test harness requires the assumption at every tick, so on the harness the two readings agree.

The evaluator is the case study's program that scores an episode against the kernel's conjuncts. The architecture model states every contract on the whole trace, and there its verdicts equal the evaluator's (\cref{cqd:sec:pillar}). Read up to the tick that decides each verdict, the model and the prefix reading differ only on excused traces (\cref{cqd:thm:prefix}). They differ on 44 of 600 random synthetic traces and on 40 with a second seed, and on no trace of another kind. \Cref{cqd:sec:case-study} reports prefix verdicts.

\section{Delegation as the Quotient}
\label{cqd:sec:delegation}

\label{cqd:sec:terms}
We write $\Csys$ for the system requirement read as a contract and $\Csib = \C_1 \comp \cdots \comp \C_k$ for the composition of the contracts of the fixed components, which we also call the slot's siblings. The \emph{slot} is the position of the missing component in the architecture. A \emph{delivery} is a candidate implementation of the slot, and it is \emph{admitted} when it satisfies its requirement on every scenario of a unit harness that mocks its neighbors (\cref{cqd:sec:pillar}). \Cref{cqd:tab:notation} gives the notation of the case study.

\begin{table}[!htb]
\centering
\footnotesize
\caption{Notation of the case study. Signals are read at a control tick; lengths in m, times in s, rates in m/s and rad/s.}
\label{cqd:tab:notation}
\begin{tabular}{@{}lp{5.9cm}lp{5.0cm}@{}}
\toprule
\textbf{Symbol} & \textbf{Meaning} & \textbf{Symbol} & \textbf{Meaning} \\
\midrule
$c_{\mathrm{true}}$ & true clearance now, robot body to nearest point of the person's body; $c_{\mathrm{true}}^{+}$ its smallest value up to the next tick & $c_0$ & required clearance \\
$c_{\mathrm{stamp}}$ & true clearance at the observation time of the report in hand & $\delta_{\mathrm{r}}$, $\delta_{\mathrm{p}}$ & tolerances of the drive's displacement and of the person's closing \\
$c_{\mathrm{rep}}$ & reported clearance, to the nearest reported person point & $v_{\mathrm{a}}$ & closing speed of the person's body \\
$\mathrm{age}$ & time since the observation in hand was measured; $\mathrm{gap}$ the time from the tick to the next control tick & $|v|$, $|w|$ & magnitudes of the commanded speed and yaw rate (\texttt{v}, \texttt{w} in the listings) \\
$e_{\mathrm{p}}$ & bound on $c_{\mathrm{rep}} - c_{\mathrm{stamp}}$ in view, the perception error & $e_{\mathrm{loc}}$, $e_{\mathrm{yaw}}$ & position and heading error of the pose estimate \\
$s_{\mathrm{r}}$ & largest commanded speed in effect over the age and the drive's 0.15~s lag before it, times the age; $s_{\mathrm{r}}^{+}$ the same with the age counted to the next tick & $g$, $y$ & distance and heading error to the dock, true or estimated ($g_{\mathrm{true}}$, $g_{\mathrm{est}}$; \texttt{g\_est} in the listings) \\
$k_{\mathrm{ext}}$ & closing of the nearest body point on the robot's position at the observation, up to the tick; $k_{\mathrm{ext}}^{+}$ up to the next tick & $\mathrm{rest}$ & the robot has been still for the settle time \\
\bottomrule
\end{tabular}
\end{table}

\subsection{The Missing Component Problem and Its Canonical Solution}
\label{cqd:sec:delegation-problem}

\begin{definition}[missing component problem]
\label{cqd:def:delegation-problem}
Given a system contract $\Csys$ and the contracts $\C_1, \ldots, \C_k$ of the fixed components, a requirement for the missing component is a saturated contract $\C$. It is sound if $\C_1 \comp \cdots \comp \C_k \comp \C \refines \Csys$. The missing component problem asks for the most permissive sound requirement, that is, a sound requirement that every sound requirement refines.
\end{definition}

Soundness is what we mean by correct by construction. Unfolded through \cref{cqd:sec:ag-contracts,cqd:def:ops}, it states that $E \cap M \cap M_1 \cap \cdots \cap M_k$ lies in the guarantee of $\Csys$ for every environment $E \in \Env(\Csys)$ and all implementations $M \in \Imp(\C)$ and $M_i \in \Imp(\C_i)$.

\begin{definition}[delegated contract]
\label{cqd:def:delegated}
The delegated contract for the missing component is the quotient $\Cq = \Csys \quotient \Csib$.
\end{definition}

\begin{corollary}[delegation is canonical]
\label{cqd:thm:canonical}
$\Cq$ is sound, and every sound requirement refines it, so $\Cq$ solves the missing component problem.
\end{corollary}

\begin{proof}
Apply \cref{cqd:prop:adjoint} with $\C_1 = \Csys$ and $\C_2 = \Csib$. Taking $\C = \Cq$ in its equivalence gives $\Csib \comp \Cq \refines \Csys$, and for a sound $\C$ the equivalence gives $\C \refines \Cq$.
\end{proof}

The corollary is the cited universal property read for delegation, and this reading is our contribution. Since $\C \refines \Cq$ implies $\Imp(\C) \subseteq \Imp(\Cq)$, every implementation that some sound requirement allows is an implementation of $\Cq$.

\subsection{Preservation Obligations}
\label{cqd:sec:preservation}

Let $(\varphi_\A, \varphi_\G)$ be the system contract, let $(\alpha_i, \gamma_i)$ be the stated pair of the $i$-th fixed component's contract, and write $\alpha = \alpha_1 \wedge \cdots \wedge \alpha_k$. By \cref{cqd:def:ops} the composition $\Csib$ has the guarantee $\gamma_{\mathrm{sib}} = \bigwedge_i (\gamma_i \vee \neg\alpha_i)$ and the assumption $\alpha \vee \neg\gamma_{\mathrm{sib}}$. Under the quotient's assumption $\neg\gamma_{\mathrm{sib}}$ is false, so
\[
\Cq \;=\; \Big(\underbrace{\varphi_\A \wedge \gamma_{\mathrm{sib}}}_{\text{assumption}},\;\;
\underbrace{(\varphi_\G \wedge \alpha)}_{\text{guarantee core}} \vee \neg(\varphi_\A \wedge \gamma_{\mathrm{sib}})\Big).
\]
The conjuncts of $\alpha$ are the preservation obligations, and a task statement that gives only the mission conjunct $\varphi_\G$ omits them. In the running example they are the three bounds on the controller's yaw rate that perception, localization and the drive assume. Each bound is a condition on the controller's own outputs, so a candidate can be checked against it with no neighbor in the loop.

\begin{definition}[mission-only requirement]
\label{cqd:def:naive}
Let $\gamma_{\mathrm{ideal}}$ be the guarantee of the fixed components taken as ideal, under which every report they hand the missing component is exact and current and every other guarantee holds as stated. The mission-only requirement for the missing component is the saturation $\Cnaive = (\varphi_\A \wedge \gamma_{\mathrm{ideal}},\; \varphi_\G \vee \neg(\varphi_\A \wedge \gamma_{\mathrm{ideal}}))$. Ideal reports meet every error bound, so $\llbracket\gamma_{\mathrm{ideal}}\rrbracket \subseteq \llbracket\gamma_1 \wedge \cdots \wedge \gamma_k\rrbracket$.
\end{definition}

\begin{theorem}[mission-only requirement is unsound]
\label{cqd:thm:naive}
Suppose a behavior $b$ exists that is system-legal ($b \in \llbracket\varphi_\A\rrbracket$), lies outside the ideal ($b \notin \llbracket\gamma_{\mathrm{ideal}}\rrbracket$), breaks the assumption of every fixed component whose guarantee it violates ($b \in \llbracket\gamma_i\rrbracket \cup \neg\llbracket\alpha_i\rrbracket$ for every $i$), and fails the mission ($b \notin \llbracket\varphi_\G\rrbracket$). Then there are an implementation $M \in \Imp(\Cnaive)$, a fixed-component implementation $M_{\mathrm{sib}} \in \Imp(\Csib)$, and a legal environment $E \in \Env(\Csys)$ whose composition $E \cap M \cap M_{\mathrm{sib}}$ contains $b$ and so violates the system guarantee. In particular $\Csib \comp \Cnaive \not\refines \Csys$.
\end{theorem}

\begin{proof}
Take $M = \llbracket\varphi_\G\rrbracket \cup \neg\llbracket\varphi_\A \wedge \gamma_{\mathrm{ideal}}\rrbracket$, the largest implementation of $\Cnaive$. Since $b \notin \llbracket\gamma_{\mathrm{ideal}}\rrbracket$, $b$ lies outside the mission-only assumption and so $b \in M$. Take $M_{\mathrm{sib}} = \bigcap_i (\llbracket\gamma_i\rrbracket \cup \neg\llbracket\alpha_i\rrbracket) = \llbracket\gamma_{\mathrm{sib}}\rrbracket$, the guarantee of $\Csib$ and so its largest implementation; $b \in M_{\mathrm{sib}}$ by the third hypothesis. Take $E = \llbracket\varphi_\A\rrbracket$. Then $b \in E \cap M \cap M_{\mathrm{sib}}$ and $b \notin \llbracket\varphi_\G\rrbracket$. The composite $M \cap M_{\mathrm{sib}}$ implements $\Csib \comp \Cnaive$, and refinement would confine its system-legal behaviors to the system guarantee, so it fails.
\end{proof}

A delegation under the mission-only requirement is therefore not correct by construction. The implementation $M$ of the proof is a set of behaviors and need not be causal (\cref{cqd:sec:realizability}). The witness of \cref{cqd:sec:res-controls} is a causal controller on the live system that keeps its own requirement in episodes in which it breaks the system's.

The hypothesis admits both mechanisms of \cref{cqd:sec:introduction}. Either $b$ keeps every fixed guarantee and leaves the ideal within their tolerances through a stale or erroneous report, or $b$ breaks a fixed component's assumption and loses its guarantee. An engineer who assumes the advertised guarantees $\gamma_{\mathrm{eff}} = \gamma_1 \wedge \cdots \wedge \gamma_k$ in place of the ideal budgets the tolerances by hand and removes the first mechanism. The proof uses only $b \notin \llbracket\gamma_{\mathrm{ideal}}\rrbracket$, so the theorem holds with $\gamma_{\mathrm{eff}}$ in place of $\gamma_{\mathrm{ideal}}$ for every behavior of the second kind. Under C-BASE the tolerances enter through the fixed components' guarantees in the quotient's assumption, and the strengthening of \cref{cqd:sec:dispatch} budgets them, as the person margin of \cref{cqd:sec:cs-dispatch} does. The preservation obligations are what the quotient adds beyond any budget of tolerances.

We call a unit-test harness any procedure that observes an implementation $M$ only through its behaviors in the ideal set $I = \llbracket\varphi_\A \wedge \gamma_{\mathrm{ideal}}\rrbracket$ and checks a formula on the traces it generates. Running $M$ against ideal mocks of the fixed components is such a procedure, and checking $\varphi_\G$ with it is the unit-test practice that the mission-only requirement formalizes.

\begin{corollary}[unit-test blindness]
\label{cqd:cor:unit-tests}
A unit-test harness generates no behavior of the kind in \cref{cqd:thm:naive}, and it gives the same verdicts to two implementations that agree on $I$. Under the hypothesis of \cref{cqd:thm:naive}, the implementation $M$ of its conclusion can be chosen to agree on $I$ with an implementation of $\Cq$, so no unit-test harness separates the two, whatever formula it checks and however many traces it runs. A unit-test harness that checks $\varphi_\G$ admits every implementation of $\Cnaive$, including those that break $\alpha$ on its traces.
\end{corollary}

\begin{proof}
The behaviors of \cref{cqd:thm:naive} satisfy $b \notin \llbracket\gamma_{\mathrm{ideal}}\rrbracket$ and so lie outside $I$, and the verdicts of a procedure that observes $M$ only on $I$ depend on $M$ only through $M \cap I$. Let $M = \llbracket\varphi_\G \wedge \alpha\rrbracket \cup \neg I$. Then $M \cap I \subseteq \llbracket\varphi_\G\rrbracket$, so $M \in \Imp(\Cnaive)$, and $b \in M$ since $b \notin I$. The proof of \cref{cqd:thm:naive} uses only these two facts about its implementation, so it holds with this $M$. The assertion $M_{\mathrm{q}} = \llbracket\varphi_\G \wedge \alpha\rrbracket$ lies inside the guarantee of $\Cq$, so $M_{\mathrm{q}} \in \Imp(\Cq)$, and $M \cap I = M_{\mathrm{q}} \cap I$. For every $M_{\mathrm{sib}} \in \Imp(\Csib)$ the composite $M_{\mathrm{q}} \cap M_{\mathrm{sib}}$ implements $\Csib \comp \Cq$, which refines $\Csys$ by \cref{cqd:thm:canonical}, so its system-legal behaviors lie in the system guarantee. Finally, every implementation $M'$ of $\Cnaive$ has $M' \cap I \subseteq \llbracket\varphi_\G\rrbracket$, whatever it does about $\alpha$, so a harness that checks $\varphi_\G$ on traces in $I$ admits it.
\end{proof}

A check of $\varphi_\G$ can reject an implementation of $\Cnaive$ only on traces outside $I$, such as those produced by mocks held at their contracts' bounds rather than at the ideal. A harness that also checks $\alpha$ rejects every implementation that breaks a preservation obligation on its traces, so the quotient makes these obligations checkable before integration.

\subsection{From Quotient to a Dispatched Requirement}
\label{cqd:sec:dispatch}

The mission conjunct of the quotient can refer to signals that the delegated component does not observe. The controller, for instance, receives the clearance to a person only as a report that is up to 0.40~s old and may overstate it by up to $e_{\mathrm{p}}$.

\begin{definition}[dispatched requirement]
\label{cqd:def:dispatched}
A dispatched requirement is a saturated contract $\Cdisp$ with $\Cdisp \refines \Cq$ whose guarantee core is written over the observables of the delegated component and the bounds declared by the fixed components' guarantees.
\end{definition}

Since $\Cdisp \refines \Cq$, \cref{cqd:prop:adjoint} gives $\Csib \comp \Cdisp \refines \Csys$, so the dispatched requirement is sound, and its implementations are among those of $\Cq$. For the clearance conjunct the strengthening rests on the following kinematic inequality.

\begin{lemma}
\label{cqd:lem:margin}
Let the environment assume $k_{\mathrm{ext}}^{+} \leq \delta_{\mathrm{p}} + v_{\mathrm{a}}\,(\mathrm{age} + \mathrm{gap})$ and that a person out of view at a tick stays at least $c_0$ away up to the next tick. Let the drive guarantee that the robot moves at most $\delta_{\mathrm{r}} + s_{\mathrm{r}}^{+}$ from its position at the observation up to the next tick, and let perception guarantee $c_{\mathrm{rep}} \leq c_{\mathrm{stamp}} + e_{\mathrm{p}}$ while the person is in view. Then at every tick at which the person is in view, $c_{\mathrm{true}}^{+} \geq c_{\mathrm{rep}} - e_{\mathrm{p}} - \delta_{\mathrm{r}} - s_{\mathrm{r}}^{+} - \delta_{\mathrm{p}} - v_{\mathrm{a}}\,(\mathrm{age} + \mathrm{gap})$. In particular the conjunct $\Alw\big(c_{\mathrm{rep}} \geq c_0 + e_{\mathrm{p}} + \delta_{\mathrm{r}} + \delta_{\mathrm{p}} + s_{\mathrm{r}}^{+} + v_{\mathrm{a}}\,(\mathrm{age} + \mathrm{gap})\big)$ implies $\Alw(c_{\mathrm{true}}^{+} \geq c_0)$ under those assumptions.
\end{lemma}

\begin{proof}
Fix a physics step from the tick up to the next tick. Let $z$ be the nearest body point at that step, $x_0$ the robot's position at the observation and $r$ the offset between a distance from the robot's position and a clearance. By the triangle inequality the distance from the robot at that step to $z$ is at least the distance from $x_0$ to $z$ less the robot's displacement since the observation. The distance from $x_0$ to $z$ is at least $c_{\mathrm{stamp}} + r - k_{\mathrm{ext}}^{+}$, by the definition of $k_{\mathrm{ext}}^{+}$. Subtracting $r$ and substituting the three bounds gives a lower bound on the clearance at that step, and hence on its smallest value $c_{\mathrm{true}}^{+}$, and the conjunct's right-hand side makes that bound at least $c_0$. At a tick with the person out of view, $c_{\mathrm{true}}^{+} \geq c_0$ is the environment's assumption.
\end{proof}

The until of the dispatched requirement keeps the speed bound and the margin of \cref{cqd:lem:margin} until a rest at which the estimated distance and heading error to the dock lie within the system's tolerances less the localization errors at rest. The bounded until is monotone in both operands, so under the fixed components' guarantees the dispatched until implies the system's, the preservation conjuncts carry over, and $\Cdisp \refines \Cq$ follows conjunct by conjunct. The kernel also checks both inclusions on the case-study instance, over sampled valuations of the tick signals and over short traces built from the lemma's worst case.

\subsection{Realizability, Evidence, and Risk}
\label{cqd:sec:realizability}
\label{cqd:sec:evidence}

The kernel checks that the dispatched requirement is consistent, that is, that some behavior meets both its assumption and its guarantee core. A consistent quotient can still be unrealizable, because the algebra describes sets of behaviors with no causality condition, not what a component can produce from what it observes. In the case study a controller that keeps $|w| \leq 0.8$~rad/s at every tick meets every preservation obligation, including the bound of 1.0~rad/s on the largest yaw rate over the 2.0~s before a stop, without anticipating its stops. The authors wrote a reference controller against the dispatched requirement before dispatch (\cref{cqd:sec:cs-protocol}), and it meets the whole dispatched requirement in the harness and on all 30 drawn episodes of the live system (\cref{cqd:sec:res-controls}).

A delivery is accepted on live episodes scored from ground truth, because by \cref{cqd:cor:unit-tests} no unit-test harness can stand in for them. The next lemma bounds what a run of satisfied episodes establishes.

\begin{lemma}[zero-failure bound]
\label{cqd:lem:binomial}
If $N$ independent episodes drawn from the operating distribution all satisfy a requirement, then with confidence $1 - \beta$ the probability of a violating episode is at most $1 - \beta^{1/N}$.
\end{lemma}

\begin{proof}
A violation probability $\riskeps$ gives zero failures in $N$ episodes with probability $(1-\riskeps)^N$. The largest $\riskeps$ for which this exceeds $\beta$ is $1 - \beta^{1/N}$.
\end{proof}

The evaluator also assigns each live failure to an owner, by a rule we adopt on top of the algebra. A failure of the system's clearance is owned by the first clause to fail, in the prefix, among those \cref{cqd:lem:margin} relies on: the environment assumption, the perception's guarantee, the drive's tracking and the person margin of $\Cdisp$. Failures at the same tick are ranked in that order, so when two fixed components fail at one tick the perception owns the failure. A failure of the speed bound alone is the delegated component's. A missed dock is ranked by clause rather than by tick. It is the environment's if its assumption failed by the deadline, else the drive's if its yaw floor failed, else localization's if its guarantee failed, else the delegated component's. A fixed component's failure is the delegated component's instead, on the preservation conjunct for that component, when the delegated component broke that component's assumption at or before the tick of the failure. If the environment assumption held and every fixed component kept its guarantee, then by refinement the delegated component violated a conjunct of $\Cdisp$, and the evaluator names which one.

\section{C-BASE: The Architecture and the Design Pipeline}
\label{cqd:sec:pillar}

\Cref{cqd:fig:orchestration} shows the parts of C-BASE. The SysML~v2 architecture model holds the robot's structure together with the delegation's requirements, verification cases and results (\cref{cqd:sec:pillar-organization}). The \emph{contract kernel}, our own JAX code, holds every contract as a pair of assertions in the fragment of \cref{cqd:sec:scoring}. It computes saturation, composition and the quotient, checks refinement on finite universes and scores traces by robustness. Its generator writes the contracts into the architecture model as requirement definitions, and the kernel reads them back through Syside's Python API, checking each constraint against the formula text it carries, so the kernel and the model hold the same contracts. The work order, the harness and the evaluator are built from the kernel's conjuncts, so the agent is scored on the formulas it receives. The orchestrator launches each step of the pipeline below. Its operators, in the case study the authors with a coding agent, declare the contracts and the strengthening onto observables.

A \emph{condition} is the artifact handed to the language model: the dispatched requirement, the mission-only requirement or a natural-language task. A \emph{form} is the kind of delivery asked for. The three forms are a whole controller, a \emph{governor}, and a selection of a governor and its parameters from a menu. A governor takes the command of the stack's route follower, which has no rule for people, and returns the command sent to the drive. A round is one call of the language model, and an \emph{arm} is one delivery run live on a set of drawn episodes and scored. A delegation then runs as follows, with steps 3 and 4 the language model's, step 6 the simulator's and every other step deterministic code.
\begin{enumerate}[leftmargin=*]
    \item \textit{Compute and render.} The kernel reads the system requirement and the fixed components' contracts from the architecture model, takes the quotient of the system contract by the composed fixed components and checks that the strengthening onto the controller's observables refines it. Each conjunct is rendered as a line of \stl{} text, such as \texttt{always (not fov or w <= 0.8)} for $|w| \leq 0.8$~rad/s while a person is in view, with a sentence on its meaning and its reason, and the margin is given term by term.
    \item \textit{Write the work order}, one JSON object with the function's interface, the operating context in words, the requirement from the architecture model, a starting file with the stack's route follower, the task and the earlier rounds with their candidates and counterexamples. Its response schema admits a candidate with its complete source or a decline with a reason.
    \item \textit{Call the design agent}, a language model served by Ollama, with the work order as its message and its decoding constrained to the response schema.
    \item \textit{Receive the delivery}, a complete Python file.
    \item \textit{Gate it in the unit harness}, a closed-loop simulation of the slot with the fixed components mocked as the condition's assumption states them. The harness runs the delivery on every scenario of a table fixed before the run and scores every conjunct as in \cref{cqd:sec:scoring}. The delivery is admitted when the assumption held and the guarantee was met in every scenario. Otherwise each failing conjunct returns its worst scenario to the next round, with the formula and the signals at its worst tick, up to the form's round limit.
    \item \textit{Run it live.} An admitted file runs unchanged in the slot of the assembled stack on the simulator, with the real perception, localization and drive and the simulator's ground truth recorded.
    \item \textit{Score and attribute.} The evaluator scores the episode by the same conjuncts under the prefix reading and attributes the first violation to its conjunct and owner (\cref{cqd:sec:evidence}).
    \item \textit{Return the evidence.} A script writes each arm's evaluation into the architecture model's result package as one result item per specification, with verdict counts, provenance and links to the requirement usage it verifies, its verification case and each violated constraint.
\end{enumerate}

\subsection{The Requirements in the Architecture Model}
\label{cqd:sec:pillar-organization}

\begin{figure}[!t]
\centering
\includegraphics{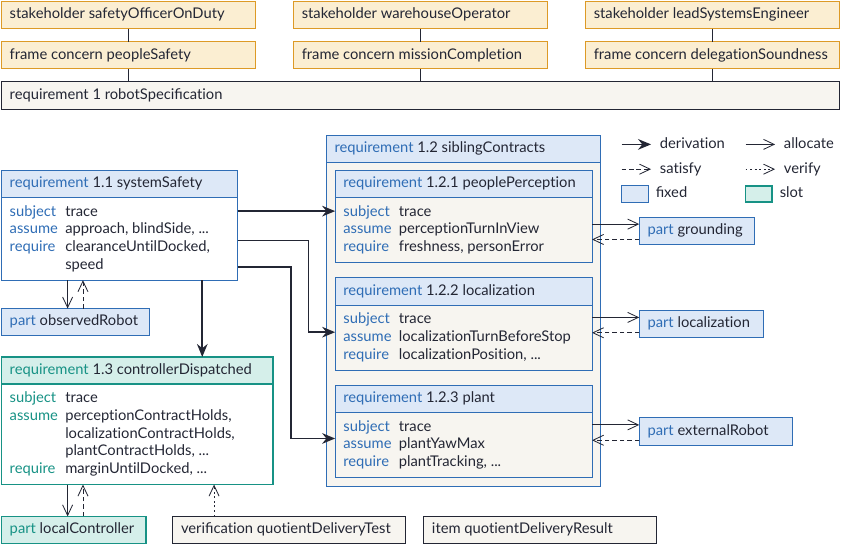}
\caption{The requirements of the case study's architecture model. Stakeholders and their concerns frame the robot's specification. Derivations run from the system requirement to the three fixed components' contracts and to the requirement dispatched to the controller slot, which strengthens the quotient (\cref{cqd:def:dispatched}). Each requirement is allocated to and satisfied by a part, and the verification case of the quotient condition verifies the dispatched requirement, with its result item holding the verdicts.}
\label{cqd:fig:requirements}
\end{figure}

The delegation's requirements sit in the model of the robot's ROS~2 stack, built on the \texttt{ros2-sysmlv2} library \citep{dameraAaC2026} and modeled after the specification's Annex~A \citep{omg2025sysml} (\cref{cqd:fig:requirements}). Each contract of \cref{cqd:tab:contracts} is one requirement definition, and each conjunct is an assumed or required constraint over the ticks of the episode trace (\cref{cqd:sec:sysml-reading}). Always becomes a \sysml{forAll} over the ticks, and the bounded until becomes an \sysml{exists} over the tick indices with a \sysml{forAll} over the earlier ones (\cref{cqd:lst:quotient-req}). Each constraint also carries its formula as text, which is the line the work order quotes. The dispatched requirement assumes the environment and, for each fixed component, the implication from that component's assumption to its guarantee. It requires the margin until docked, the speed rule and the three preservation conjuncts, and its numbered usage carries metadata that names the kernel computation it came from.

\begin{lstlisting}[language=SysMLv2, float=tp, caption={The dispatched requirement as the kernel's generator writes it, and its numbered usage, abridged.}, label={cqd:lst:quotient-req}]
requirement def DispatchedControllerRequirement {
    subject trace : EpisodeTrace;
    // cut: the doc, the assumed constraints and the four other required constraints
    require constraint marginUntilDocked {
        attribute formulaText : String = "(v <= 0.5 and c_rep >= 0.74 + s_r_gap + 1.5 * age + 1.5 * gap) until[0, 90 s] (rest and g_est <= 0.05 and y_est <= 0.12)";
        (1..size(trace.samples))->exists {in k : Integer;
            trace.samples#(k).elapsedS <= 90.0
            and trace.samples#(k).atRest
            and trace.samples#(k).estimatedGoalDistanceM <= 0.05
            and trace.samples#(k).estimatedGoalYawErrorRad <= 0.12
            and (1..size(trace.samples))->forAll {in j : Integer;
                j >= k or (trace.samples#(j).commandSpeedMps <= 0.5
                    and trace.samples#(j).reportedClearanceM >= 0.74
                        + 1.5 * trace.samples#(j).observationAgeS
                        + 1.5 * trace.samples#(j).gapToNextTickS
                        + trace.samples#(j).robotStaleMotionToNextTickM)
            }
        }
    }
}
requirement <'1.3'> controllerDispatched : DispatchedControllerRequirement {
    @DerivedByQuotient {
        formula = "the kernel quotient of SystemRequirement by the composition of PerceptionContract, LocalizationContract and PlantContract, strengthened to observable signals ...";
    }
}
\end{lstlisting}

\subsection{Satisfaction, Verification and Evidence}
\label{cqd:sec:pillar-verification}

Each component's part specializes both its structural definition and the episode trace, so a recorded episode is the part's trace, and the part satisfies the requirement usage allocated to it. Each condition has a verification case whose objective verifies the condition's requirement usage on the recorded episode, and further cases verify single conjuncts of the fixed components' contracts. A result item counts the verdicts of one arm over its episodes and names the reading they were scored under, the evaluator's prefix reading or the whole trace (\cref{cqd:sec:scoring}).

Syside's validator checks the model files with the library at zero errors and zero warnings, resolving every name and type. On hand-built traces its compiler evaluates every constraint to the value of its formula text read tick by tick, and on random traces to the verdict of the study's evaluator. Copies of the model seeded with one defect each, among them the drive's contract satisfied by the controller, are each caught by these checks or by a check that resolves every link as a model element. Every value the script writes into a result item equals its source. The supplementary material gives the counts.

\section{Case Study: A Delegated Robot Controller}
\label{cqd:sec:case-study}

The case study asks three questions. Does a delivery admitted under the mission-only requirement break the system requirement live because it ignores a tolerance in a neighbor's guarantee, while a delivery admitted under the derived requirement keeps it? Does the derived requirement refuse a delivery that breaks a neighbor's assumption, which the mission-only requirement admits? What does a language model deliver under each requirement and under a natural-language task, and how do its deliveries fare live?

\subsection{System, Mission and Contracts}
\label{cqd:sec:cs-system}
\label{cqd:sec:cs-contracts}

The robot is a Carter warehouse robot in NVIDIA Isaac Sim \citep{isaacSim2026} with a Python navigation stack on ROS~2 \citep{macenski2022robot}, whose perception (a camera detector grounded on the lidar cloud), localization and drive are fixed and whose local controller is the slot. At 20~Hz the slot reads the planned path, the pose estimate, the person points with their observation age and the lidar scan, and publishes a speed and a yaw rate. It wraps the delivery in fixed code that brakes on a missing or stale observation, and what that code commands is scored as the delivery's.

The robot must drive a 12~m route across open floor and rest at the dock at its end within 0.10~m and 0.15~rad by 90~s, never faster than 0.5~m/s. A worker starting 1.5~m beside the route crosses it at a drawn point and speed (0.5 to 0.85~m/s), straight or angled 0.35~rad toward the robot, triggered so that a robot that never yields meets the worker. This seeded population is the operating distribution of \cref{cqd:lem:binomial}. The slot sees the oldest observation within the freshness bound, so the live system runs at the observation age that the perception contract allows, and ground truth is the simulator's record of the robot's pose and the worker's root and 55 joints. Clearances are measured from the robot's body, a 0.30~m disc, and the robot is at rest after 1.5~s with its commands and true speed below 0.02~m/s and 0.02~rad/s. Each bound of \cref{cqd:tab:contracts} comes from its mechanism, and a bound read from live records lies above the largest value seen, as the table shows for the worker's closing and the drive's yaw following.

\providecolor{envC}{RGB}{221,154,38}
\providecolor{envT}{RGB}{251,238,210}
\providecolor{fixedC}{RGB}{47,109,181}
\providecolor{fixedT}{RGB}{221,232,247}
\begin{table}[!htb]
\centering
\footnotesize
\caption{The contracts of the case study, one clause per row, with the mechanism that sets each bound. Notation as in \cref{cqd:tab:notation}; lengths in m, times in s, rates in m/s and rad/s. The architecture model names $\C_{\mathrm{P}}$, $\C_{\mathrm{L}}$ and $\C_{\mathrm{D}}$ \sysml{PerceptionContract}, \sysml{LocalizationContract} and \sysml{PlantContract}; they are the fixed components' contracts $\C_i$ of \cref{cqd:sec:delegation}.}
\label{cqd:tab:contracts}
\setlength{\tabcolsep}{4pt}
\begin{tabular}{@{}>{\raggedright\arraybackslash}p{2.2cm}>{\raggedright\arraybackslash}p{7.7cm}>{\raggedright\arraybackslash}p{3.6cm}@{}}
\toprule
\textbf{Clause} & \textbf{Bound} & \textbf{Set by} \\
\midrule
\rowcolor{envT} \multicolumn{3}{@{}c@{}}{\textcolor{envC}{\rule[0.6ex]{2em}{0.5pt}}\enspace\textcolor{envC}{\textbf{System $\Csys$ assumes}}\enspace\textcolor{envC}{\rule[0.6ex]{2em}{0.5pt}}} \\
approach & $k_{\mathrm{ext}} \leq 0.02 + 1.5\,\mathrm{age}$ and $k_{\mathrm{ext}}^{+} \leq 0.02 + 1.5\,(\mathrm{age} + \mathrm{gap})$ & largest limb closing measured, 1.385~m/s \\
blind side & nobody within 0.40 out of view, now and to the next tick & the camera's field of view \\
frames & a scan and a people record every 0.5 & sensor rates \\
walk & crosses once, within 0.72 of its segment, back at most 0.72 & the drawn walks \\
route & the route 1.4 from mapped surfaces & the warehouse layout \\
\rowcolor{fixedT} \multicolumn{3}{@{}c@{}}{\textcolor{fixedC}{\rule[0.6ex]{2em}{0.5pt}}\enspace\textcolor{fixedC}{\textbf{System $\Csys$ guarantees}}\enspace\textcolor{fixedC}{\rule[0.6ex]{2em}{0.5pt}}} \\
speed & $|v| \leq 0.5$ at every tick & the mission \\
clearance to dock & $(|v| \leq 0.5 \wedge c_{\mathrm{true}}^{+} \geq 0.40) \Until_{[0,90]} (\mathrm{rest} \wedge g_{\mathrm{true}} \leq 0.10 \wedge y_{\mathrm{true}} \leq 0.15)$ & the mission \\
\midrule
\rowcolor{envT} \multicolumn{3}{@{}c@{}}{\textcolor{envC}{\rule[0.6ex]{2em}{0.5pt}}\enspace\textcolor{envC}{\textbf{Perception $\C_{\mathrm{P}}$ assumes}}\enspace\textcolor{envC}{\rule[0.6ex]{2em}{0.5pt}}} \\
yaw in view & $|w| \leq 0.8$ while a person is in view & image-to-cloud skew \\
\rowcolor{fixedT} \multicolumn{3}{@{}c@{}}{\textcolor{fixedC}{\rule[0.6ex]{2em}{0.5pt}}\enspace\textcolor{fixedC}{\textbf{Perception $\C_{\mathrm{P}}$ guarantees}}\enspace\textcolor{fixedC}{\rule[0.6ex]{2em}{0.5pt}}} \\
age & $\mathrm{age} \leq 0.40$ & admission age and cloud period \\
error & $c_{\mathrm{rep}} \leq c_{\mathrm{stamp}} + 0.30$ in view & the detector and the skew \\
\midrule
\rowcolor{envT} \multicolumn{3}{@{}c@{}}{\textcolor{envC}{\rule[0.6ex]{2em}{0.5pt}}\enspace\textcolor{envC}{\textbf{Localization $\C_{\mathrm{L}}$ assumes}}\enspace\textcolor{envC}{\rule[0.6ex]{2em}{0.5pt}}} \\
turn before stop & $|w| \leq 1.0$ over the 2.0 before every stop & rest heading after turns, measured \\
\rowcolor{fixedT} \multicolumn{3}{@{}c@{}}{\textcolor{fixedC}{\rule[0.6ex]{2em}{0.5pt}}\enspace\textcolor{fixedC}{\textbf{Localization $\C_{\mathrm{L}}$ guarantees}}\enspace\textcolor{fixedC}{\rule[0.6ex]{2em}{0.5pt}}} \\
position at rest & $e_{\mathrm{loc}} \leq 0.05$ & one map cell \\
heading at rest & $e_{\mathrm{yaw}} \leq 0.03$ & standstill search reach \\
settling & still, below 0.01~m/s, within 1.5 of stillness & timing of the standstill correction \\
position moving & $e_{\mathrm{loc}} \leq 0.50$ & declared bound \\
\midrule
\rowcolor{envT} \multicolumn{3}{@{}c@{}}{\textcolor{envC}{\rule[0.6ex]{2em}{0.5pt}}\enspace\textcolor{envC}{\textbf{Drive $\C_{\mathrm{D}}$ assumes}}\enspace\textcolor{envC}{\rule[0.6ex]{2em}{0.5pt}}} \\
yaw rate & $|w| \leq 1.0$ & the turn-rate saturation \\
\rowcolor{fixedT} \multicolumn{3}{@{}c@{}}{\textcolor{fixedC}{\rule[0.6ex]{2em}{0.5pt}}\enspace\textcolor{fixedC}{\textbf{Drive $\C_{\mathrm{D}}$ guarantees}}\enspace\textcolor{fixedC}{\rule[0.6ex]{2em}{0.5pt}}} \\
tracking & displacement at most $s_{\mathrm{r}} + 0.02$ over the age, $s_{\mathrm{r}}^{+} + 0.02$ to the next tick & breakaway and command lag \\
yaw floor & a command $\geq 0.25$ held 0.5 below 0.2~m/s turns the body $\geq 0.0625$ & the dead band \\
yaw following & true yaw rate within 0.3 of the last 0.6 of commands & largest live excess, 0.214~rad/s, plus 0.05~rad/s \\
\bottomrule
\end{tabular}
\end{table}

\subsection{Delegation Artifacts and Harness}
\label{cqd:sec:cs-dispatch}
\label{cqd:sec:cs-harness}

The dispatched requirement $\Cdisp$ keeps the assumption of $\Cq$ and writes its guarantee over the slot's inputs. Its until keeps the speed at most 0.5~m/s and the person margin $c_{\mathrm{rep}} \geq 0.74 + s_{\mathrm{r}}^{+} + 1.5\,(\mathrm{age} + \mathrm{gap})$~m of \cref{cqd:lem:margin} until a rest within 90~s. At that rest the estimated distance and heading error to the dock are within 0.05~m and 0.12~rad, the system's dock tolerances less the localization errors at rest. Its preservation conjuncts are the three yaw-rate bounds that the fixed components assume (\cref{cqd:tab:contracts}). The kernel checks $\Cdisp \refines \Cq$ and $\Csib \comp \Cdisp \refines \Csys$, with the contracts read from the architecture model, on finite universes built from the lemma's worst case, and both inclusions hold. Neither holds for the mission-only requirement, which fails on the guarantee side as well as on the assumption side. The first inclusion also fails when any margin term, the preservation conjuncts, the speed rule, the dock margin or the rest in the dock is dropped from $\Cdisp$, and on a universe that breaks the lemma by 0.05~m. The supplement gives the universes and the counterexample counts.

The mission-only requirement renders $\Cnaive$ onto the slot's inputs with the fixed components ideal. It assumes the environment, the drive's guarantee, an observation age of zero, exact person points and an exact pose, and requires the speed rule and a reported clearance of 0.40~m until an estimated rest within 0.10~m and 0.15~rad of the dock. The natural-language task states in words, with no formula, the system requirement and every fixed component's bounds with the turning limit each one assumes. These are the facts from which the dispatched requirement is derived, and the task leaves the margin and the dock tolerances on the observables to the designer.

The harness simulates a straight 10~m route ending at the dock, on 482 scenarios fixed before every run. They cover walkers crossing straight or angled toward the robot over a grid of crossing points, sides, speeds and trigger offsets, the same walks with the person points reported closer than they are, and two scenarios without a person. The harness grew to these scenarios from a first version of 98, the straight walks and the empty scenarios, on which the witness of \cref{cqd:sec:cs-protocol} and the contract variants of \cref{cqd:sec:res-family} were run. Under the quotient and natural-language conditions the mocks sit at the contract bounds: the observation is 0.40~s old, the person points are pushed 0.30~m away along their rays and are not reported out of view, and the drive saturates at 1.15~rad/s. The pose is off by 0.05~m at rest and by 0.45~m while moving, the largest error seen live within the bound of 0.50~m. Under the mission-only condition the mocks are exact, as that requirement assumes (\cref{cqd:cor:unit-tests}). Each condition is scored by its own requirement, the natural-language one by $\Csys$ on the mocked truth.

\subsection{Contract Calibration and Experimental Protocol}
\label{cqd:sec:cs-siblings}
\label{cqd:sec:cs-protocol}

Two controllers written by the authors serve as controls. The \emph{reference} follows the route turning at most 0.8~rad/s and settles in the dock. It holds the robot while a reported person point lies ahead within a band around the route, widened by its pose error, so a worker about to cross holds the robot and one who has crossed lets it pass. The robot's position across the route is known within 0.05~m at rest but only within 0.50~m while moving, so the reference carries the pose of its last rest forward on the drive's unicycle model, in a memory that the slot passes back at every tick. It also stops when the dispatched margin is about to fail. The \emph{witness}, written before any run to the letter of the mission-only requirement, brakes on the reported clearance with a hand-chosen allowance of 0.20~m, docks on the estimate while moving and turns at the stack's 1.9~rad/s. It is the control for the harness's discrimination and for the live evaluator, and the mission-only arm proper is the agent's delivery under that condition.

Every bound of \cref{cqd:tab:contracts} has a verification case that prints its threshold before its result. The fixed components were held to their guarantees on calibration episodes, driven by the reference and kept out of every arm, and each one found outside its contract was repaired, or its contract was, before any delivery was judged. Calibration changed four bounds and repaired one component, the perception's camera projection. The localization's heading bound, for instance, holds only at rest, because the estimate lags the body by up to 0.229~rad in a turn in place. The supplement lists every correction with its finding.

The agent is \texttt{qwen3.8:\allowbreak 27b-q8\_0}, a model with 27 billion parameters at 8-bit weights, served by Ollama on the simulator's workstation and sampled at temperature 0.4 with a seed recorded per round. It is asked for the three forms of \cref{cqd:sec:pillar} in that order, with up to six rounds each, and the first admitted delivery ends the condition. The menu of the third form holds 384 governors, of which 288 keep no memory and 96 are band governors, the reference's design with six of its parameters varied. Each request carries a record of the designs that earlier rounds and earlier sequences of its condition tried, with their outcomes, a view of the route and one from the robot's camera, and after a tested round a plot of its counterexample. The supplement gives the remaining settings and every round.

Each admitted delivery runs live on the 30 standard scenarios, drawn once and shared with the controls. An episode in which the slot braked for missed control deadlines more than 200 times ran on an overloaded host, and it is run again on a quiet host, the re-run taking its place. The live evaluator reads the worker's approach on the walker's joints, which swing ahead of its root and change identity between ticks, and subtracts a reading allowance of 0.05~m, beside the 0.02~m person tolerance. The harness walker is a disc whose extent moves with its centre, so the harness under-predicts the live clearance.

\subsection{Admission Under Each Requirement}
\label{cqd:sec:res-agent}
\label{cqd:sec:results}
The derived requirement admits only safe designs, and the mission-only requirement does not. Of the 384 governors in the menu, the dispatched requirement admits 12, and every one of them keeps the system requirement with the fixed components at their contract bounds, as soundness promises. The mission-only requirement admits 112, and 88 of them break the system requirement at those bounds. Its harness scores a controller against ideal neighbors, so it cannot tell the 24 safe entries it admits from the 88 unsafe ones.

The agent's selections show the difference (\cref{cqd:tab:agent}). Under the derived requirement it selected one of the 12. Under the mission-only requirement its first selection happened to fall among the 24 safe entries, and three of twenty further selections with fresh seeds fell among the 88. We took the first of those three, a band governor 1.25~m wide that the harness breaks in 8 of its 482 scenarios at the bounds, to the live system as the unsafe delivery. Under the natural-language task it selected a governor that keeps the system requirement at the bounds. In no condition did the agent deliver a whole controller or governor that its harness admitted, in six rounds per form, and each condition then handed it the menu. The rounds and the counterexamples returned to the agent are in the supplement.
\begin{table}[!t]
\centering
\footnotesize
\caption{The deliveries and the controls. Admission is by the harness of each delivery's own condition. A design is safe at the bounds when it keeps the system requirement with the fixed components at their contract bounds. The live columns count the episodes in which the system requirement held on the truth, over the 30 standard scenarios and at the setting of \cref{cqd:sec:res-controls} where the neighbors work at the edge of their contracts, and the last column gives the closest the robot came to the worker in any episode.}
\label{cqd:tab:agent}
\label{cqd:tab:controls}
\setlength{\tabcolsep}{4pt}
\begin{tabular}{@{}>{\raggedright\arraybackslash}p{2.7cm}>{\raggedright\arraybackslash}p{2.5cm}>{\raggedright\arraybackslash}p{1.7cm}>{\raggedright\arraybackslash}p{1.5cm}>{\raggedright\arraybackslash}p{1.5cm}>{\raggedright\arraybackslash}p{1.5cm}@{}}
\toprule
\textbf{Delivery} & \textbf{Admitted by its own harness} & \textbf{Safe at the bounds} & \textbf{Live, 30 standard} & \textbf{Live, at the edge} & \textbf{Closest} \\
\midrule
agent, derived requirement & yes & yes & 30 of 30 & 4 of 4 & 1.42~m \\
agent, mission-only & yes & yes & 30 of 30 & not run & 1.57~m \\
agent, mission-only, resampled & yes & no, breaks 8 of 482 & 30 of 30 & 16 of 22 & 0.14~m \\
agent, natural language & yes, by its own scoring & yes & 30 of 30 & not run & 1.22~m \\
reference, hand-written & yes, under all three & yes & 30 of 30 & not run & 1.56~m \\
witness, written to the mission-only letter & yes, by the first mission-only harness & no & 0 of 30 & not run & contact in 29 \\
\bottomrule
\end{tabular}
\end{table}
\subsection{Live Results}
\label{cqd:sec:res-controls}
Every delivery kept the worker safe on the 30 standard scenarios, the unsafe one included, so ordinary testing does not separate them. The design selected under the derived requirement kept the system requirement and the dispatched requirement in all 30 episodes and came no closer than 1.45~m to the worker, which by \cref{cqd:lem:binomial} bounds the probability of a violating episode by 0.095 at confidence 0.95. The other three selections kept the system requirement in all 30 as well. The first mission-only selection missed the dispatched requirement in 1 episode, and the resampled and the natural-language selections in 9 each, on the dock tolerance and the person margin, obligations the mission-only requirement never states.

\begin{figure}[!t]
\centering
\includegraphics{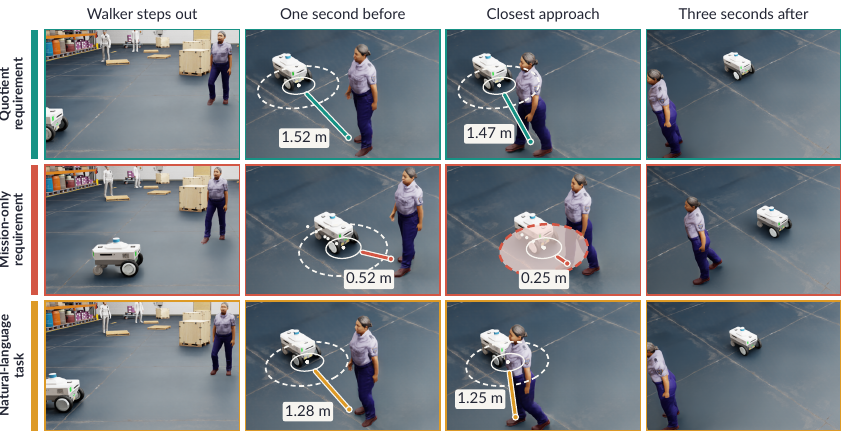}
\caption{The three deliveries live on one draw at the setting of \cref{cqd:sec:res-controls}, one row per delivery and four frames matched by the walker's progress, from a run recorded for this figure and outside the counts of the text. In the two middle frames the evaluator's geometry is drawn into the camera view: the robot's body disc, the required 0.40~m clearance as the dashed circle, the walker's nearest body point with the clearance to it, and the robot's positions over the preceding 5~s. The deliveries selected under the derived requirement and under the natural-language task stopped before the walker stepped out, 1.47~m and 1.25~m from her at the closest. The mission-only delivery drove on for three more seconds and stopped 1.27~m further along, in her path; at the closest approach her foot lies inside the circle, 0.25~m in the frame and 0.22~m between frames, the one red mark of the figure.}
\label{cqd:fig:live}
\end{figure}

Where the neighbors work at the edge of their contracts, the unsafe delivery came within 0.14~m of the worker against the required 0.40~m, and the delivery selected under the derived requirement stayed more than 1.4~m away. \Cref{cqd:fig:live} shows the three deliveries on one draw at this setting. The harness finds such a setting inside the fixed components' contracts, with the localization at its moving error bound, the reported person point pushed toward its error bound and the worker crossing the route at 0.6~m/s. There the unsafe delivery broke the system's clearance inside the environment assumption in 6 of 22 episodes and docked in every one, and the derived requirement's delivery kept both requirements in every episode on the same draws. The setting was searched for, so 6 of 22 is the rate there and not over ordinary scenarios. Each breach is a behavior of the kind \cref{cqd:thm:naive} describes. The environment assumption held, every fixed component kept its guarantee without behaving ideally, and the delivery had passed the mission-only harness.

The witness, which the first version of the mission-only harness admitted and the derived requirement refuses, broke the clearance in all 30 standard episodes, with contact in 29.
\subsection{Preservation Obligations, Contract Variants and the Perception Slot}
\label{cqd:sec:res-family}
A controller that breaks the localization's assumption passes the mission-only harness, and the derived requirement refuses it. The localization promises its rest accuracy only after a turn under 1.0~rad/s. A controller that spins at 1.2~rad/s and holds still before the dock is admitted by the mission-only harness in every scenario and refused by the dispatched requirement in every scenario, on that obligation. Live, in 9 of 9 episodes, the spin engaged the localization's declared lock at the rest after it, the estimate docked while the robot stood 0.26~m from the dock, and the mission reported success on a false dock. The reference on the same draws never engaged the lock and docked by the truth every time. The reference itself was caught the same way. As first written it turned at 1.9~rad/s, which the mission-only requirement admitted and the dispatched requirement refused, and it now turns at most 0.8~rad/s.

Change one neighbor's contract and the derived requirement follows it, still admitting only safe designs, while the mission-only requirement does not move (\cref{cqd:tab:variants}). On the first harness and the 288 governors without a memory, we relaxed the perception's freshness bound to 0.80~s, tightened it to 0.20~s, and coarsened the localization's rest error to 0.20~m, one change at a time. The quotient tracks the set of safe governors exactly in every variant. The dispatched requirement admits a subset of it that shrinks to nothing when the perception slows, which is the cost of its strengthening. The mission-only requirement admits the same 120 governors whatever the neighbors do, 96 of them unsafe at the declared bounds. On the full harness it admits 112 of the 384 governors of the menu instead (\cref{cqd:sec:res-agent}). The refinement order among the variants' requirements is in the supplement.
\begin{table}[!htb]
\centering
\footnotesize
\caption{The 288 governors without a memory admitted as one fixed component's contract changes, on the first harness of 98 scenarios. GT counts the governors that keep the system requirement with the fixed components at the variant's bounds, Q those the quotient $\Cq$ admits and D those the dispatched requirement admits, both at those bounds, and M those the mission-only requirement admits against ideal neighbors, which do not change.}
\label{cqd:tab:variants}
\begin{tabular}{@{}llllll@{}}
\toprule
\textbf{Variant} & \textbf{Changed bound} & \textbf{GT} & \textbf{Q} & \textbf{D} & \textbf{M} \\
\midrule
as declared & age 0.40~s, rest error 0.05~m & 24 & 24 & 12 & 120 \\
slower perception & age 0.80~s & 12 & 12 & 0 & 120 \\
faster perception & age 0.20~s & 24 & 24 & 12 & 120 \\
coarser map & rest error 0.20~m & 0 & 0 & 0 & 120 \\
\bottomrule
\end{tabular}
\end{table}

Delegating the perception instead of the controller gives the same result. With the reference as the fixed controller, the quotient makes freshness within 0.40~s and a person error within 0.30~m the perception slot's obligations. A slot written to the letter of the mission-only requirement delivers reports older than that bound under the declared overload and never docks. The agent's grounding code written against the slot's quotient kept the system requirement and the dispatched requirement in all 30 live episodes, with its observation at most 0.40~s old at every control tick. It was admitted after its first live run exposed a dropped lidar cloud, which the harness then gained as a declared scenario. The delivery rounds and the slot's dropped records are in the supplement.

\section{Conclusion}
\label{cqd:sec:conclusion}

In requirements-driven MBSE, delegating a component when the rest of the architecture is fixed poses the missing component problem, and its solution is the quotient of the system requirement by the fixed components' contracts. In SysML~v2 that quotient is again a requirement, the most permissive sound one, so the designer keeps the most freedom while the architecture stays correct by construction. C-BASE computes it with a contract kernel, keeps it in the architecture model, hands it to a language model as a work order, admits deliveries in a harness and scores them live. On a simulated warehouse robot, the language model's selection under the quotient kept the system requirement on every live episode, and so did the perception grounding it wrote against that slot's quotient. After a dropped cloud became a harness scenario, the model rewrote the grounding and the harness admitted it again. A reference written against the quotient kept the system requirement on every live episode, while the witness, the literal mission-only controller admitted by the mission-only harness it was written against, brought the robot below the required clearance on every live episode. A controller that broke the localization's assumption by turning fast before its dock left the robot more than 0.25~m from the dock on all nine episodes of that arm, while its own estimate reported the dock reached. On the same draws the reference never engaged the localization's lock.

\backmatter

\section*{Significance and Practitioner Points}

For practitioners, the requirement handed to a supplier, a team, or a generative agent for one component of a system can be derived from the system requirement and the requirements of the other components by one algebraic operation. The result is an ordinary SysML~v2 requirement that sits in the model with its derivation, allocation, satisfaction and verification links. The derived requirement states, as explicit constraints, what the other components assume of the delegated one, which is exactly what task descriptions omit and what integration testing discovers late. For researchers, this article grounds the contract quotient in the requirement semantics of SysML~v2 and reports a controlled comparison of quotient-derived, mission-only and natural-language delegation to a language model on a robot navigation stack in simulation.

\section*{Acknowledgements}
The authors thank Sensmetry for providing academic licenses for the Syside Automator and Modeler tools.

\section*{Conflict of Interest}
The authors declare no conflict of interest.

\section*{Data Availability Statement}
The SysML~v2 model, the contract kernel, and the campaign records described in this article will be made available on publication.

\bibliography{bibliography/references}
\end{document}